\documentclass[nofootinbib,twocolumn,amssymb,nobibnotes,aps,pre,superscriptaddress]{revtex4-1}

\usepackage{amsmath}
\usepackage{amsfonts}
\usepackage{amssymb}
\usepackage{graphicx}
\usepackage{hyperref}
\usepackage{longtable}
\usepackage{dcolumn}
\usepackage{bm}
\usepackage{epstopdf}
\usepackage{color}

\begin{document}

\title{Dynamics of passive and active membrane tubes}

\author{Sami C.\ Al-Izzi}
\affiliation{School of Physics \& EMBL-Australia node in Single Molecule Science, University of New South Wales, Sydney, Australia}
\affiliation{Department of Mathematics, University of Warwick, Coventry CV4 7AL, UK}
\affiliation{Institut Curie, PSL Research University, CNRS, Physical Chemistry Curie, F-75005, Paris, France}
\affiliation{Sorbonne Universit\'{e}, CNRS, UMR 168, F-75005, Paris, France}

\author{Pierre Sens}
\affiliation{Institut Curie, PSL Research University, CNRS, Physical Chemistry Curie, F-75005, Paris, France}
\affiliation{Sorbonne Universit\'{e}, CNRS, UMR 168, F-75005, Paris, France}

\author{Matthew S.\ Turner}
\affiliation{Department of Physics, University of Warwick, Coventry CV4 7AL, UK}
\affiliation{Centre for Complexity Science, University of Warwick, Coventry CV4 7AL, UK}
\affiliation{Department of Chemical Engineering, University of Kyoto, Kyoto 615-8510, Japan}

\author{Shigeyuki Komura}
\affiliation{Department of Chemistry, Graduate School of Science, Tokyo Metropolitan University, Tokyo 192-0397, Japan}

\date{\today}

\begin{abstract}
Utilising Onsager's variational formulation, we derive dynamical equations for the 
relaxation of a fluid membrane tube in the limit of small deformation,
 allowing for a contrast of solvent viscosity across the membrane and variations in surface tension due to membrane incompressibility.
We compute the relaxation rates, recovering known results in the case of purely axis-symmetric perturbations and making new predictions for higher order (azimuthal) $m$-modes. We analyse the long and short wavelength limits of these modes by making use of various asymptotic arguments. 
We incorporate stochastic terms to our dynamical equations suitable to describe both passive thermal forces and non-equilibrium 
active forces. We derive expressions for the fluctuation amplitudes, an effective temperature associated with active fluctuations, and the power spectral density for both the thermal and active fluctuations.
We discuss an experimental assay that might enable measurement of these fluctuations to infer the properties of the active noise. Finally we discuss our results in the context of active membranes more generally and give an overview of some open questions in the field.
\end{abstract}

\maketitle

\section{Introduction}
\label{sec:introdcution}

Membrane tubes, formed by bilayers of phospholipid molecules, are structures that are ubiquitous in cells. 
They are vital to the function of many organelles including the peripheral Endoplasmic Reticulum 
(ER)~\cite{nixon-abell_increased_2016}. So-called membrane nanotubes have been identified more recently and implicated as a
pathway in inter-cellular signalling~\cite{abounit_wiring_2012}. Membrane tubes can be formed from a patch of membrane by the action of a 
localised normal force on the membrane, {\em e.g.}, from molecular motors such as myosin or kinesin, or from the formation of a coat of intrinsically curved proteins on the membrane~\cite{derenyi_formation_2002,cuvelier_coalescence_2005,yamada_catch-bond_2014}.

From a statistical mechanics perspective, there already exists a significant body of work on the thermal fluctuations of membrane tubes~\cite{fournier_critical_2007,komura_fluctuations_1992}.  A striking prediction from these theories is that the bending modes of the tube are critical   in the long wavelength limit, meaning that fluctuations are predicted to diverge at the linear level. Anharmonic terms in the free energy then control the excess area and associated length fluctuations~\cite{fournier_critical_2007}. These studies have gained contemporary relevance with the development of fluorescence microscopy and optical tweezers 
techniques able to infer the power spectral density of fluctuations on tubes pulled from Giant 
Unilamelar Vesicles (GUVs)~\cite{valentino_fluctuations_2016}. 
Such studies may also have some relevance to the statistical mechanics of tubular 
networks~\cite{tlusty_microemulsion_2000,tlusty_topology_2000} and on the length fluctuations of tubes held 
by a fixed force~\cite{barbetta_fluctuations_2009}.

Work on the dynamics of membrane tubes has focused on the simplified axisymmetric case, in particular the dynamics 
of the pearling instability of membrane 
tubes~\cite{bar-ziv_instability_1994,boedec_pearling_2014,nelson_dynamical_1995, gurin_dynamic_1996}, and the 
dynamics of tether pulling from a GUV or cell~\cite{evans_hidden_1994,nassoy_nanofluidics_2008,brochard-wyart_hydrodynamic_2006}. 
A further area of study is that of particle lateral mobility within the membrane~\cite{DanielsTurner_2007,henle_hydrodynamics_2010,rahimi_curved_2013}. 
These examples demonstrate how the curved geometry of the membrane tube can lead to rich physical phenomena, in the 
form of visco-elastic couplings~\cite{rahimi_curved_2013} and non-Newtonian rheological 
behaviour~\cite{brochard-wyart_hydrodynamic_2006,evans_hidden_1994}.

Recently there has been an increased interest in quantifying the dynamics and fluctuations of membrane tubes in a biological setting~\cite{nixon-abell_increased_2016} and in understanding the transport dynamics of cargo within membrane tubes~\cite{abounit_wiring_2012,marbach_transport_2018,holcman_ER_flow_2018}. Such scenarios are, in general, driven far from equilibrium by active forces from cytoskeletal interactions~\cite{turlier_equilibrium_2016} or proteins such as proton pumps changing conformation when consuming ATP \cite{ramaswamy_nonequilibrium_2000,al-izzi_hydro-osmotic_2018}. In order to have a complete physical understanding of such processes one needs to develop a theory of active membrane tubes.

In this paper, we will focus on the dynamics of membrane tubes, deriving equations of motion from Onsager's variational 
principle in the manner of Refs.~\cite{fournier_hydrodynamics_2015,sachin_krishnan_relaxation_2016,sachin_krishnan_thermal_2018},
and analysing the relaxation dynamics in Fourier space. 
We then consider the case where stochastic forces act on the membrane and derive the statistical properties of the shape 
undulations, in particular focusing on the case where active noise dominates. 
Here the term active refers to a noise term which breaks detailed balance. 
Such active membrane systems have been studied extensively for the case of flat membranes~\cite{prost_shape_1996,ramaswamy_nonequilibrium_2000,gov_membrane_2004} and 
spherical vesicles~\cite{sachin_krishnan_thermal_2018,turlier_equilibrium_2016}. 
These descriptions also have relevance for experiments incorporating active proteins into
GUVs~\cite{manneville_active_2001} and in the analysis of red blood cell flicker~\cite{turlier_equilibrium_2016,gov_red_2005,garcia_direct_2015}.

In Sec.~\ref{sec:Onsager} we derive hydrodynamic equations of motion for a membrane tube utilising the Osager's variational principle. We decompose these equations of motion in Fourier space, allowing the equations to be reduced to an overdamped equation for the perturbations in the radial displacement of the tube surface.
The relaxation dynamics of the radial displacement are discussed in Sec.~\ref{sec:relaxationDynamics}. The onset of a pearling instability is discussed in Sec.~\ref{sec:pesrlingInstability}. In Sec.~\ref{sec:fluctuations} we analyse the fluctuations of membrane tubes due to stochastic forces of two types; the first corresponding to thermal fluctuations and the second corresponding to a simple form of active noise that breaks detailed balance.
We derive the fluctuation spectra for these ``active'' tubes and calculate a wave length-dependent effective temperature of such fluctuations, to be compared to thermal fluctuations. Finally in Sec.~\ref{sec:discussion} we discuss possible ways to quantify the parameters in our active fluctuations model from experiment, the relation of our work to previous studies 
and some open problems in the study of membrane tubes and active membranes more generally.

\section{Membrane tube dynamics}
\label{sec:Onsager}

Here we introduce the geometry required to build our model, and the notation we will use. We then derive equations of motion for the relaxation dynamics of the membrane tube in the linear response regime.
\subsection{Geometry}

\begin{figure}
\includegraphics[width=0.45 \textwidth]{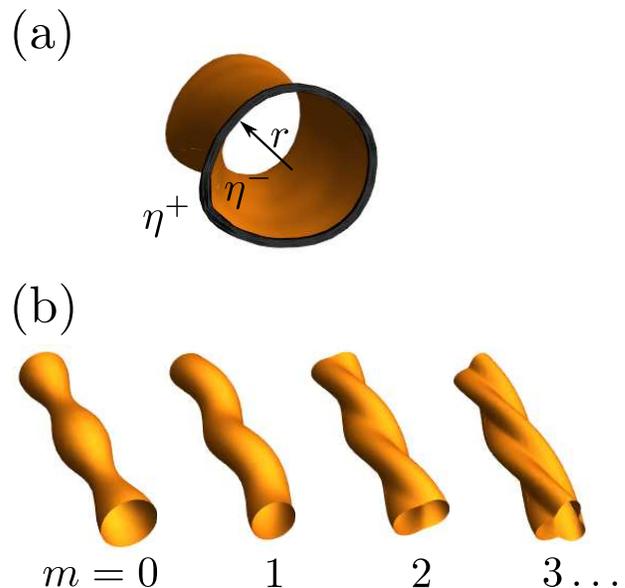}
\caption{(a) Cross section of a membrane tube with time-dependent undulation of its radial position $r=r_0 \left[1+u\left(\theta,z,t\right)\right]$ about the 
equilibrium radius $r_0$.
Here $\eta^+$ and $\eta^-$ are the viscosity of the exterior and interior ambient fluid, respectively. 
(b) Surface plots of the Fourier decompositions, $u=\sum_{q,m}u_{qm}e^{i q z+i m\theta}$ for $q r_0=1$.} 
\label{fig:schematic}
\end{figure}

We treat the membrane as a two-dimensional manifold, $\mathcal{S} \subset \mathbb{R}^3$. 
Vectors in the ambient space will be denoted
$\vec{x} \in \mathbb{R}^3$ and vectors in the tangent bundle to 
the manifold as $\boldsymbol{x}\in \mathcal{T}\left(\mathcal{S}\right)$.
We parametrise the manifold, $\mathcal{S}$, with the vector 
$\vec{r}=(r \cos \theta, r \sin \theta,z )$ where 
$r(\theta,z,t) = r_0 \left[1 + u\left(\theta,z,t\right)\right]$, see Fig.~\ref{fig:schematic}(a). 
We will consider the small deformation limit where $u\ll 1$. 
Local tangent vectors can be induced on the surface by taking derivatives with respect to $\theta$ and $z$, giving 
$\vec{e}_\theta=\partial_\theta \vec{r}$ and $\vec{e}_z=\partial_z\vec{r}$, respectively.
A complete triad can be defined by $\{\vec{e}_\theta,\vec{e}_z,\vec{n}\}$ where 
$\vec{n}=(\vec{e}_\theta\times \vec{e}_z)/ |\vec{e}_\theta\times \vec{e}_z|$ 
is the normal vector to the surface. 
The metric and second fundamental (bilinear) forms are then defined as $g=g_{ij}\mathrm{d}X^i \mathrm{d}X^j=
\vec{e}_i \cdot \vec{e}_j \mathrm{d}X^i \mathrm{d}X^j$ and $b=b_{ij}\mathrm{d}X^{i}\mathrm{d}X^{j}$ where 
$b_{ij}=\vec{n}\cdot\partial_{j}\vec{e}_{i}$ and $\mathrm{d}X^i $ are the coordinate basis of the cotangent bundle. 
By raising the indices of the second fundamental form with the metric and taking the trace and determinant, 
we define the mean curvature, $H=b_i{}^i/2$, and Gaussian curvature, $K=\det b_i{}^j$.

The membrane is assumed to behave as a fluid in the tangential direction and we define a vector flow field of the lipids in the membrane as 
$\boldsymbol{v}\in\Gamma\left(\mathcal{T}\left(\mathcal{S}\right)\right)$, where $\Gamma\left(\mathcal{T}\left(\mathcal{S}\right)\right)$ is a section of the tangent bundle. 

\subsection{Elastic free energy}

The elastic free energy of an incompressible fluid membrane can be written using the Helfrich-Canham-Willmore
energy~\cite{helfrich_elastic_1973,canham_minimum_1970,willmore_note_1965}
\begin{equation}
\mathcal{F}_{\text{el}} = \int_{\mathcal{S}}\mathrm{d}A\left[ \sigma +\frac{\kappa}{2}(2H )^2\right],
\end{equation}
where  $\sigma$ is the surface tension, $\kappa$ the bending rigidity and the area element is given by 
$\mathrm{d}A=\sqrt{|g|} \mathrm{d}\theta\mathrm{d}z$. 
We have integrated out the contribution from the Gaussian curvature and saddle splay modulus by assuming no changes 
in topology and treat the tube as having infinite length. 
We also neglect spontaneous curvature (set it equal to zero), assuming that we are considering a symmetric bilayer.

The area element and the mean curvature squared are given, up to second order, by
\begin{equation}
\mathrm{d}A \approx r_0
\left[1+ u +\frac{1}{2}\left(r_0^2 \partial_z^2u + \partial_\theta^2u\right)\right] \mathrm{d}\theta\mathrm{d}z,
\end{equation}
\begin{align}
H^2 & \approx \frac{1}{r_0^2}\bigg[\frac{1}{2} -r_0^2 \partial_z^2u+\partial_\theta^2 u +u 
\nonumber \\
& + \frac{1}{2}\Big\{2 u (r_0^2 \partial_z^2 u+3\partial_\theta^2 u)-r_0^2 (\partial_z u)^2
\nonumber \\
&+(r_0^2 \partial_z^2 u+\partial_\theta u)^2+ (\partial_\theta u)^2+3 u^2\Big\} \bigg].
\end{align}

If we take the full elastic free energy $\mathcal{F}=\mathcal{F}_{\text{el}}-\Delta P\int \mathrm{d}V$, where 
$\Delta P = P^--P^+$ is the hydrostatic pressure difference across the membrane, then the ground state $r=r_0$ must satisfy the modified Laplace equation
\begin{equation}\label{eq:laplace}
\frac{\sigma}{r_0}-\frac{\kappa}{2r_0^3}-\Delta P=0,
\end{equation}
in order to minimise the elastic free energy.

For the Onsager's formulation of membrane dynamics, we need to know the rate of change of the free energy. 
This is given by
\begin{align}
\dot{\mathcal{F}} & = \int\mathrm{d}\theta\mathrm{d}z \bigg[\frac{\kappa}{r_0^2} \bigg(\frac{3}{2}u +\frac{1}{2}r_0^2\partial_z^2u +r_0^4\partial_z^4u +\frac{5}{2}\partial_\theta^2u 
\nonumber \\ 
& + 2 r_0^2\partial_\theta^2\partial_z^2u + \partial_\theta^4u \bigg) - \sigma (u + r_0^2\partial_z^2u +\partial_\theta^2u )\bigg]r_0 \dot{u},
\label{eq:freeEnergyRate}
\end{align}
where a dot $(\>\dot{}\>)$ indicates a time derivative and we have made use of Eq.~(\ref{eq:laplace}), or equivalently the constraint that 
total volume is preserved $\int \mathrm{d}\theta\mathrm{d}z\, u = -\int \mathrm{d}\theta\mathrm{d}z\, u^2/2$. 
Note that $\dot{\mathcal{F}}$ is a functional only of the normal velocity $v_n=r_0 \dot{u}+\mathcal{O}(u^2)$ 
and not the tangential components of membrane velocity $\boldsymbol{v}$. This is also true for arbitrarily large shape perturbations \cite{arroyo_relaxation_2009}.

\subsection{Dissipation and constraints}

We will consider only the dissipation due to the ambient fluid as this is the dominant dissipative mechanism at large 
length-scales~\cite{seifert_hydrodynamics_1994}. 
At the scale of cell membranes ($10$~nm--$100$~$\mu$m), viscous dissipation dominates the dynamics of the fluid.
Hence we neglect the contribution from inertia and assume zero Reynolds number~\cite{happel_low_1983}.

We define the velocity in the ambient fluid as $\vec{V}=V^{\alpha}\vec{e}_{\alpha}$, where we use Greek indices 
to denote coordinates in $\mathbb{R}^3$, and summation over repeated indices is assumed. 
The dissipation functional for the bulk fluid is given by~\cite{landau_fluid_2013}
\begin{equation}\label{eq:dissipation}
\mathcal{P}^{\pm} = \int_{\mathcal{V}^\pm} \mathrm{d}V^\pm\,
\eta^\pm D_{\alpha\beta}^{\pm}D^{\alpha\beta \pm},
\end{equation}
where $\eta^\pm$ is the viscosity and $\mathrm{d}V^{\pm}$ is the volume element of the exterior ($\mathcal{V}^{+}$) and interior ($\mathcal{V}^{-}$) regions respectively, as shown in 
Fig.~\ref{fig:schematic}(a). 
Moreover, $D_{\alpha\beta}^{\pm} = (\nabla_{\alpha}V^{\pm}_{\beta}+\nabla_{\beta}V^{\pm}_{\alpha})/2$ 
is the rate-of-strain tensor where $\nabla_\alpha$ is the ambient covariant derivative in $\mathbb{R}^3$.

Our system has several constraints which, in the Onsager's formulation, will be imposed using Lagrange
multipliers~\cite{doi_soft_2013}. 
Firstly, the membrane and ambient fluid are incompressible so must satisfy the following conditions
\begin{equation}
\nabla_{\alpha}V^{\alpha \pm}=0,
\end{equation}
for the bulk fluid and
\begin{equation}\label{eq:membraneIncompressibility}
\nabla_iv^i -2 v_n H=0,
\end{equation}
for the membrane. 
Further constrains come in the form of no-slip and no-permeation boundary conditions on the bulk fluid at the membrane:
\begin{equation}
(V^{\alpha \pm}|_{r_0})^i=v^i,~~~~~V^{r \pm}|_{r_0}=r_0 \dot{u}
\label{boundarycond}
\end{equation}
where the Latin indices denote the projection of the velocities in $\mathbb{R}^3$ onto the tangent basis of the membrane. 

\subsection{Rayleighian and equations of motion}

To derive the full equations of motion using the Onsager's formulation, we must first write down the 
Rayleighian~\citep{doi_onsagers_2011,landau_fluid_2013,fournier_hydrodynamics_2015,sachin_krishnan_relaxation_2016}. 
 The full Rayleighian for the system is found by taking the sum of the rate-of-change of free energy for the system,
Eq.~(\ref{eq:freeEnergyRate}), and the energy dissipations, Eq.~(\ref{eq:dissipation}), and adding in the 
constraints on the system using Lagrange multipliers. 
This formulation is equivalent to Onsager's kinetic equation with reciprocal coefficients, but recast as a variational 
formalism~\citep{onsager_reciprocal_1931,onsager_reciprocal_1931-1,doi_onsagers_2011}.

Thus our Rayleighian reads 
\begin{widetext}
\begin{align}\label{eq:Rayleighian}
\mathcal{R} & = \mathcal{P}^+ +\mathcal{P}^- +\dot{\mathcal{F}} 
+ \int_{\mathcal{S}} \mathrm{d}A \, \zeta ( \nabla_iv^i+\dot{u}) 
- \int_{\mathcal{V}^+} \mathrm{d}V^+ \, P^+\nabla_\alpha V^{\alpha +} 
- \int_{\mathcal{V}^-} \mathrm{d}V^- \, P^-\nabla_\alpha V^{\alpha -}
\nonumber \\ 
& + \int_{\mathcal{S}} \mathrm{d}A \,
\Big[ \mu^+_i\left[(V^{\alpha +}|_{r_0})^i-v^i\right]
+ \mu^-_i\left[(V^{\alpha -}|_{r_0})^i-v^i\right] 
+ \lambda^+\left(V^{r +}|_{r_0}-r_0 \dot{u} \right) 
+ \lambda^-\left(V^{r -}|_{r_0}-r_0 \dot{u} \right)\Big],
\end{align}
\end{widetext}
where $\zeta$, $P^\pm$, $\mu^\pm_i$ and $\lambda^\pm$ are the Lagrange multipliers imposing our constraints. 
Note that we choose the sign for $P^\pm$ and $\zeta$ so that they correspond to pressure and surface tension variation, respectively.

We now proceed to use Onsager's principle and minimise the Rayleighian to find the equations of motion for the membrane~\cite{doi_soft_2013}. 
Taking variations of Eq.~(\ref{eq:Rayleighian}) with respect to $V^{\alpha \pm}|_{r_0}$ yields
\begin{equation}
\mp \eta^\pm D^\pm_{r i}|_{r_0}-\mu^\pm_i=0,
\end{equation}
\begin{equation}
\mp \eta^\pm D^\pm_{rr}|_{r_0} \pm P^\pm -\lambda^\pm = 0,
\end{equation}
showing that $\mu^\pm_i$ and $\lambda^\pm$ correspond to the traction forces acting on the membrane.

Extremising with respect to $v^i$ gives
\begin{equation}
\nabla_i\zeta -\mu^+_i -\mu^-_i=0.
\end{equation}
By eliminating the Lagrange multipliers, we further have 
\begin{equation}
\nabla_i\zeta + \eta^+D^+_{ri}-\eta^-D^-_{ri}=0,
\end{equation}
which is simply tangential force balance on the membrane.

Taking variations with respect to $r_0\dot{u}$ and eliminating $\lambda^\pm$, we obtain normal force balance 
on the membrane 
\begin{align}\label{eq:normalForceBalanceVariation}
& \frac{\kappa}{r_0^3} \left(\frac{3}{2}u +\frac{1}{2}r_0^2\partial_z^2u +r_0^4\partial_z^4u 
+\frac{5}{2}\partial_\theta^2u + 2 r_0^2\partial_\theta^2\partial_z^2u + \partial_\theta^4u \right) 
\nonumber \\ 
& - \frac{\sigma}{r_0} (u + r_0^2\partial_z^2u +\partial_\theta^2u ) + \frac{\zeta}{r_0}
\nonumber \\ 
& -\eta^+D_{rr}^++P^+ +\eta^-D_{rr}^- -P^- =0.
\end{align}

Varying with respect to $\zeta$ simply gives the membrane incompressibility condition, 
Eq.~(\ref{eq:membraneIncompressibility}).
Varying with respect to $V_{\alpha}^{\pm}$ and $P^\pm$ gives the usual Stokes equations
and incompressibility condition, respectively, 
\begin{equation}
\eta^\pm\nabla^2V^{\alpha \pm} = \nabla^{\alpha}P^\pm,~~~~~
\nabla_{\alpha}V^{\alpha \pm} = 0.
\end{equation}

\subsection{Fourier mode decomposition}

Next we solve the equations for the bulk fluid and calculate the traction forces on the membrane. 
Here we make use of the known solution to the Stokes equations in cylindrical coordinates 
given by Ref.~\cite{happel_low_1983} 
\begin{equation}
\vec{V}^{\pm} = \nabla \phi^\pm + \nabla \times\left(\psi^\pm \vec{e}_z\right) 
+ r\partial_r \nabla\xi^\pm +\partial_z\xi^\pm \vec{e}_z,
\end{equation}
\begin{equation}
P^{\pm} = -2 \eta^\pm \partial^2_z\xi^\pm,
\end{equation}
where $(\phi^\pm,\psi^\pm,\xi^\pm)$ are scalar functions that each satisfy the Laplace equation. 
We decompose these functions in Fourier space in $\theta$ and $z$ in terms of the coordinate systems harmonic basis
\begin{equation}
\left(\begin{matrix}
\phi^\pm\\[1.0ex]
\psi^\pm\\[1.0ex]
\xi^\pm
\end{matrix}\right) = \sum_{q,m}\left(\begin{matrix}
\Phi^\pm_{qm}\\[1.0ex]
\Psi^\pm_{qm}\\[1.0ex]
\Xi^\pm_{qm}
\end{matrix}\right) \Pi^\pm_{qm}(r)e^{i q z + i m \theta},
\end{equation}
with
\begin{equation}
\Pi^{\pm}_{qm}(r)=\begin{cases}
\Pi^+_{qm}(r) = K_{m}(qr),
\\
\Pi^-_{qm}(r) = I_{m}(qr).
\end{cases}
\end{equation}
In the above, $I_{m}(qr)$ and $K_{m}(qr)$ are modified Bessel functions of the first and second kind, respectively.

We now introduce the Fourier transform as defined by 
$f(\theta,z)=\sum_{q,m} f_{qm}e^{i q z +i m \theta}$. 
The form of surfaces given by the $m$-mode perturbations is shown in Fig.~\ref{fig:schematic}(b). 
Applying the boundary conditions on the bulk flow in Fourier space allows us to find $\Phi^\pm_{qm}$, 
$\Psi^\pm_{qm}$, and $\Xi^\pm_{qm}$ in terms of the variables $\dot{u}$, $v^\theta$, and $v^z$. 
The boundary conditions in Fourier space read
\begin{widetext}
\begin{equation}\label{eq:boundaryCondition}
\left(\begin{matrix}
r_0\dot{u}_{qm}\\[1.0ex]
v^\theta_{qm}\\[1.0ex]
v^z_{qm}
\end{matrix}\right) = \left(\begin{matrix}
\Phi^\pm_{qm}\partial_r\Pi^\pm_{qm} + (i m/r) \Psi^\pm_{qm} \Pi^\pm_{qm} 
+ \Xi^\pm_{qm}\partial_r^2\Pi^\pm_{qm}
\\[1.0ex]
(i m/r) \Phi^\pm_{qm}\Pi^\pm_{qm} - \Psi^\pm_{qm} \partial_r\Pi^\pm_{qm} +i m\Xi^\pm_{qm}\left(\partial_r \Pi^\pm_{qm} - \Pi^\pm_{qm}/r \right)
\\[1.0ex]
i q \Phi^\pm_{qm} \Pi^\pm_{qm} +ir q \Xi^\pm_{qm} 
\left(\partial_r \Pi^\pm_{qm}+ \Pi^\pm_{qm}/r\right) \end{matrix}\right)_{r=r_0},
\end{equation}
\end{widetext}
where the right hand side is evaluated at $r=r_0$.
Then we can make use of the continuity equation to eliminate 
$v^\theta_{qm} = r_0(q v^z_{qm} -i \dot{u}_{qm})/m$ and find $\Phi^\pm_{qm}$, $\Psi^\pm_{qm}$, and $\Xi^\pm_{qm}$ in terms of $\dot{u}_{qm}$ and $v^z_{qm}$, which are given in Appendix \ref{app:coefficients}.

In Fourier space, the components of the tangential force balance equation read
\begin{align}
&\frac{i m}{r_0} \zeta_{qm} +\eta^+\left[r \partial_{r} (V^{\theta +}_{qm}/r) +\frac{i m}{r}V^{r +}_{qm}\right]_{r=r_0}
\nonumber \\
&- \eta^-\left[r \partial_{r} (V^{\theta -}_{qm}/r) +\frac{i m}{r}V^{r -}_{qm}\right]_{r=r_0} = 0,
\end{align}
\begin{align}
&i q \zeta_{qm} +\eta^+\left(i q V^{r +}_{qm} +\partial_{r} V^{z +}_{qm}\right)_{r=r_0}
\nonumber \\
&- \eta^-\left(i q V^{r -}_{qm} +\partial_{r} V^{z -}_{qm}\right)_{r=r_0} = 0
\end{align}
where the bulk velocity terms can be expressed using $\Phi^\pm_{qm},\Psi^\pm_{qm}$ and $\Xi^\pm_{qm}$, and they are thus just functions of $\dot{u}_{qm}$ 
and $v^z_{qm}$. Solving for $v^z_{qm}$ and $\zeta_{qm}$ allows us to write 
$\Phi^\pm_{qm}$, $\Psi^\pm_{qm}$, and $\Xi^\pm_{qm}$ in terms of $\dot{u}_{qm}$.

Finally, by substituting back into Eq.~(\ref{eq:normalForceBalanceVariation}), we obtain the linear response equation for the shape in Fourier space as 
\begin{equation}\label{eq:dyanmicalEquationModes}
B(Q,m) \dot{u}_{qm} = -A(Q,m)u_{qm},
\end{equation}
where $Q=qr_0$ and 
\begin{align}
A(Q,m) & =\left(Q^2+m^2\right)^2 - 
\frac{1}{2}Q^2 -\frac{5}{2}m^2 +\frac{3}{2} 
\nonumber \\
& -\Sigma\left(1-Q^2-m^2\right),
\end{align}
\begin{align}\label{eq:linearResponseFriction}
&B(Q,m) =\frac{r_0^3}{\kappa}
\Bigg[\frac{\zeta_{qm}}{r_0} + 
\Bigg(P^{+}_{qm} - \eta\left(\chi+1\right)\partial_{r}V^{r +}_{qm}
\nonumber \\  
&+ \eta\left(1-\chi\right)\partial_{r} V^{r -}_{qm}
 -P^{-}_{qm}\Bigg)_{r=r_0}\Bigg]\left(\dot{u}_{qm}\right)^{-1}.
\end{align}
In the above, we have introduced the notations $\Sigma=\sigma r_0^2/\kappa$, $\eta=\eta^+ +\eta^-$, 
and $\chi = (\eta^+-\eta^-)/\eta$. The function $B(Q,m)$ is sometimes referred to as the Fourier transform of the inverse of the Oseen kernel~\cite{shlomovitz_membrane-mediated_2011}. In the absence of hydrostatic pressure difference across the tube membrane, \textit{i.e.}, $\Delta P=0$, the tube radius is set by a balance between bending and tension stresses, and we have $\Sigma=1/2$~\cite{derenyi_formation_2002}. Note that $\zeta_{qm}$, $P^{\pm}_{qm}$ and $V^{r,\theta \pm}_{qm}$ are, after solving the tangential force balance equations, simply proportional to $\dot{u}_{qm}$. Hence by dividing by $\dot{u}_{qm}$ in Eq.~(\ref{eq:linearResponseFriction}) we get the friction coefficient at the linear response level.

The exact form of $B$ is in general too complex to write down except for the $m=0$ case for which it is
\begin{align}
B(Q,0) & = \frac{\eta r_0^3}{\kappa} (1+Q^2) \bigg[\frac{\left(1-\chi\right)I_1^2}{2Q I_1I_0-
Q^2\left(I_0^2-I_1^2\right)}
\nonumber \\ 
& +\frac{\left(\chi+1\right) K_1^2}{2QK_1K_2-Q^2\left(K_1^2-K_0^2\right)}\bigg],
\end{align}
where all the modified Bessel functions are evaluated at $r=r_0$. 
For larger values of $m$, we evaluate $B$ numerically using Mathematica (Wolfram Research, Champaign, IL).
It is interesting to note that, compared with the spherical case, the $m$ and $Q$ modes are not 
constrained by the other as they are in the case of spherical
harmonics~\cite{sachin_krishnan_relaxation_2016}.

We note that the behaviour of $B$ in the limit of $Q\ll 1$ can be computed for the $m=0$ mode as 
\begin{align}
B(Q,0) \approx \frac{\eta r_0^3}{\kappa Q^2}\left[2(1-\chi) +\frac{\chi+1}{2\log(2/Q)-1-2\gamma}\right],
\end{align}
where $\gamma\approx0.577$ is the Euler-Mascheroni constant.

\section{Relaxation dynamics}\label{sec:relaxationDynamics}

For some initial condition at $t=0$, the solution to Eq.~(\ref{eq:dyanmicalEquationModes}) is given by 
$u_{qm}(t)=u_{qm}(0)\exp[-\lambda(Q,m) t]$, where $\lambda(Q,m)=A(Q,m)/B(Q,m)$ describes the 
rate at which an undulation in the tubes radius decays back to the ground state.
In this section, we analyse the form of $\lambda$ to understand the stability and relaxation dynamics of the Fourier modes 
in the shape of the membrane tube.

\begin{figure}
\includegraphics[width = 0.45\textwidth]{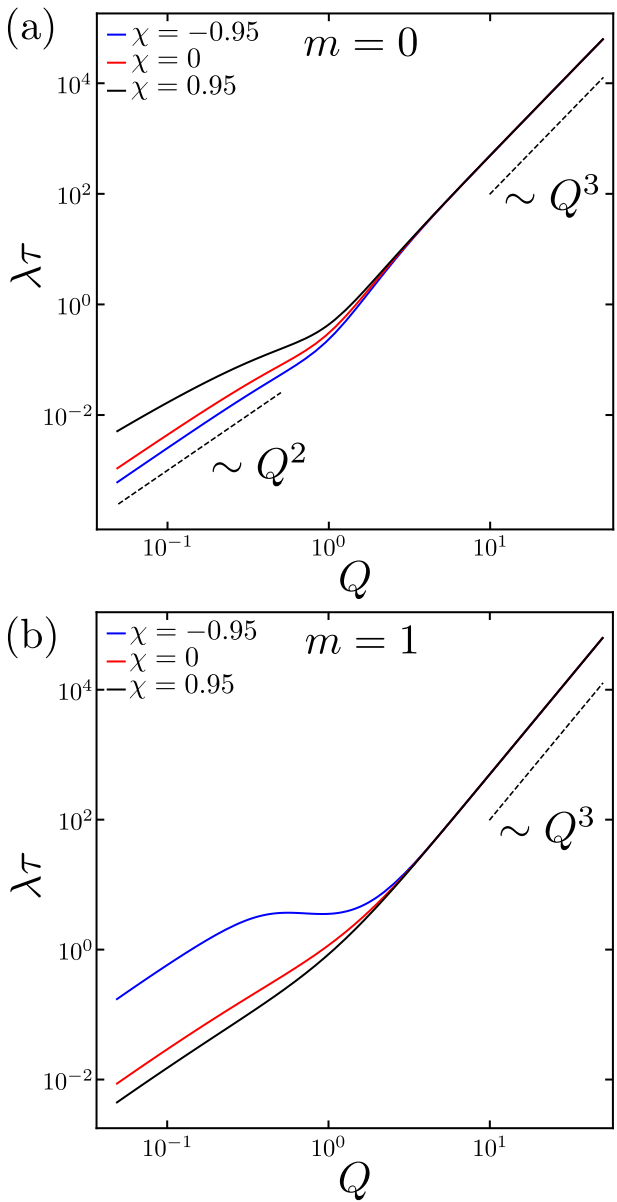}
\caption{Plot of dimensionless decay rate $\lambda\tau$ as a function of dimensionless wavenumber $Q$ 
for the modes $u_{qm}$ when (a) $m=0$ and (b) $m=1$ for varying relative viscosity $\chi=(\eta^+-\eta^-)/\eta$.
We keep $\eta$ fixed as the rate $\lambda$ has been non-dimensionalized by the viscous time associated 
with the total viscosity $\tau=\eta r_0^3/\kappa$. 
The dimensionless surface tension is given by $\Sigma=1/2$ such that the ground state has no hydrostatic 
pressure discontinuity.}
\label{fig:relaxationRates1}
\end{figure}

\begin{figure}
\includegraphics[width = 0.45\textwidth]{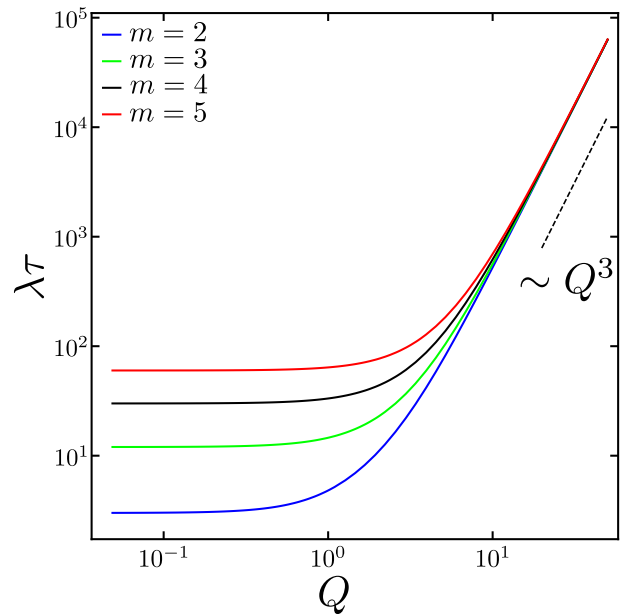}
\caption{Plot of dimensionless decay rate $\lambda\tau$ as a function of dimensionless wavenumber 
$Q$ for the modes $u_{qm}$ when $m=2,3,4,5$. We keep $\eta$ fixed as the rate $\lambda$ has been non-dimensionalized by the viscous time associated 
with the total viscosity $\tau=\eta r_0^3/\kappa$.
The dimensionless surface tension is given by $\Sigma=1/2$ such that the ground state has no hydrostatic 
pressure discontinuity.}
\label{fig:relaxationRates2}
\end{figure}

The decay rates $\lambda(Q,m)$ are plotted as a function of dimensionless wavenumber $Q=r_0 q$ 
in Figs.~\ref{fig:relaxationRates1} and \ref{fig:relaxationRates2}. 
Throughout, we fix the total viscosity $\eta=\eta^++\eta^-$ and vary the relative viscosity $\chi=(\eta^+-\eta^-)/\eta$. The 
plots are shown in units associated with the time-scale of the total viscosity given by $\tau=\eta r_0^3/\kappa$. As the dynamics remains unchanged on changing the sign of both $m$ and $Q$, we will restrict our discussion to $m,Q\geq 0$.

Let us first discuss the $m=0$ modes.
For the value of $\Sigma=1/2$, corresponding to an equilibrium ground state with no net pressure, 
the relaxation rate is given by
\begin{align}\label{eq:relaxationRate_m0}
\lambda_0(Q) & = \frac{A(Q,0)}{B(Q,0)} 
\nonumber \\
& = \left[Q^4- \frac{1}{2}Q^2 +\frac{3}{2}-\Sigma(1-Q^2)\right] 
\nonumber \\
&\times \bigg[\frac{\eta r_0^3 (1 +Q^2)}{\kappa}
\bigg(\frac{\left(1-\chi\right)I_1^2}{2Q I_1I_0-Q^2\left(I_0^2-I_1^2\right)}
\nonumber \\ 
&+\frac{\left(\chi+1\right) K_1^2}{2QK_1K_2-Q^2\left(K_1^2-K_0^2\right)}\bigg)\bigg]^{-1},
\end{align}
which is positive and the undulations are always stable.
The above expression gives the scaling behaviour $\lambda_0\sim Q^2$ in the small $Q$ regime, as 
shown in Fig.~\ref{fig:relaxationRates1}(a).
As expected in the large $Q$ limit, the scaling behaviour coincides with that of a flat membrane where 
$\lambda_0\sim Q^{3}$, and all relative viscosities converge to a universal relaxation rate.
This is a consequence of the approximate symmetry between the interior and exterior at such small length scales.

The behaviour of $\lambda$ for $m =1$ is evaluated numerically and shown in Fig.~\ref{fig:relaxationRates1}(b). 
For large values of $Q$, the scaling behaviour is again like that of a flat membrane for the same reason as the $m=0$ modes.
However, at small values of $Q$, some interesting phenomena is encountered, which is strongly dependent on the relative viscosity. 
As $Q\to 0$, the external dissipation due to the tube being dragged through the fluid dominates the relaxation rate.

When $|m|=1$, the internal motion simply corresponds to locally translating the cross section of the tube. In the long wavelength limit these gradients in velocity become small and thus the dissipation associated with the interior fluid decreases. 
In this long wavelength limit, the tube behaves like an elastic rod immersed in a viscous fluid in terms of its relaxation, and tends slowly towards 
$\lambda \sim -\left(\gamma +\log Q\right)Q^2$ as the interior dissipation becomes less dominant. 
This scaling behaviour for a continuous Zimm model of an elastic rod under tension is discussed in 
Appendix~\ref{app:zim}.

Notice that the case $\chi=-1$ corresponds to $\eta^+=0$.
Hence, in the long wavelength limit, there is essentially no friction, and the $m=1$ mode relaxation rate diverges 
as $Q\to 0$.
The crossover between interior to exterior dominant dissipation means that, in the limit of $\eta^+\ll\eta^-$, 
the relaxation rate can be non-monotonic in $Q$.
Hence $\lambda$ first decreases and then increases at intermediate $Q$ before being screened by the exterior viscosity at long wavelengths, as seen in the case $\chi=-0.95$ in Fig.~\ref{fig:relaxationRates1}(b).

For higher modes of $|m|\geq 2$, the dissipation is dominated in the long-wavelength regime by the gradients 
in velocity coming from the cross-sectional deformations of the tube. 
Thus, as $Q$ decreases, the relaxation rate becomes constant, as shown in Fig.~\ref{fig:relaxationRates2}. 
This constant increases with $m$ because each successive mode costs more bending energy to excite, so 
will relax faster.
In the high $Q$ limit, the relaxation rate scales like that of a flat membrane with $\lambda \sim Q^3$ for all $m$.
We plot the $|m|\geq 2$ modes only for $\chi=0$ because changing $\chi$ does not noticeably alter the relaxation rates for these modes as the higher $m$ modes behave like a flat membrane, and hence their relaxation depends only on the constant total viscosity, $\eta$.

\section{Pearling instability}\label{sec:pesrlingInstability}

For the $m=0$ mode, there is an instability when the tube is placed under high surface tension~\cite{boedec_pearling_2014}. 
The growth rate or dispersion relation of such an instability is given by the negative of Eq.~(\ref{eq:relaxationRate_m0}). 
The threshold for the instability at $Q=0$ is given by $\Sigma=3/2$, which corresponds to the point when 
$A(0,0)$ changes sign~\cite{nelson_dynamical_1995,gurin_dynamic_1996}.

\begin{figure}
\includegraphics[width=0.45\textwidth]{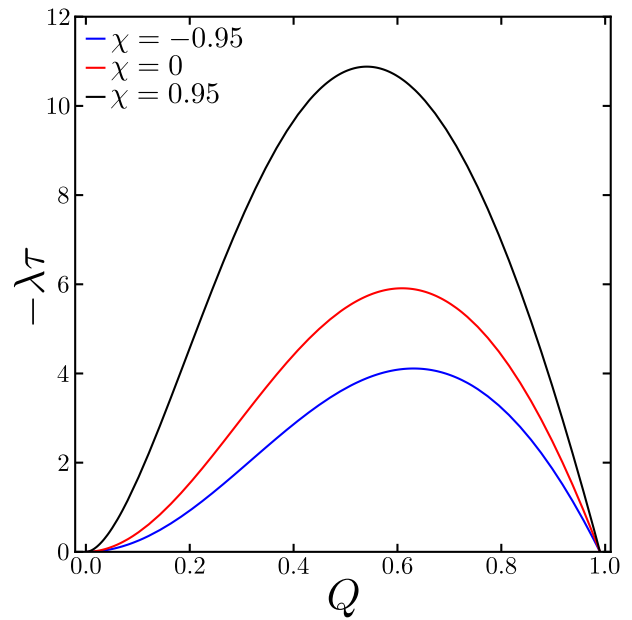}
\caption{Plot of dimensionless growth rate $-\lambda\tau$ for the pearling instability 
as a function of dimensionless $Q$ for varying values of relative viscosity $\chi$. We keep $\eta$ fixed as the rate $\lambda$ has been non-dimensionalized by the viscous time associated 
with the total viscosity $\tau=\eta r_0^3/\kappa$.
The dimensionless surface tension is set here to $\Sigma=100$.}
\label{fig:PearlingInstability}
\end{figure}

This instability is analogous to the Rayleigh-Plateau instability in a column of fluid~\cite{tomotika_instability_1935,rayleigh_xvi._1892}, 
where forces arising from the interface surface tension act to minimise the total interface area-to-volume ratio, 
and thus the fluid breaks up into spherical droplets. 
Similar forces arise in the case of membrane tubes although these are counteracted by the presence of membrane bending 
rigidity, $\kappa$. 
The exact form of this instability growth rate was found previously in Ref.~\cite{boedec_pearling_2014}, where only 
axisymmetric perturbations were considered, and was shown to coincide
with earlier works when variations in surface tension were 
neglected~\cite{nelson_dynamical_1995,gurin_dynamic_1996,powers_dynamics_2010}.

For large surface tension $\Sigma$ and similar values of viscosity ($\chi\approx 0$), the wavenumber with the maximum of 
the growth rate, $\text{argmax}_{Q}\left\{-\lambda_0\left(Q\right)\right\}$, is a monotonic function of $\Sigma$ 
which rapidly approaches $Q \approx 0.6$ \cite{nelson_dynamical_1995,boedec_pearling_2014}.
The growth rate, $-\lambda_0$, is plotted in Fig.~\ref{fig:PearlingInstability} for different values of the relative 
viscosity $\chi$.
Note that short wavelength perturbations, $Q\geq 1$, are always stable as the surface tension terms in $A(Q,0)$ 
are always positive for $Q\geq 1$.

\section{Fluctuations of membrane tubes}
\label{sec:fluctuations}

We now consider the relaxation dynamics of the tube under thermal and active fluctuations. 
This is given by adding thermal and active forces to Eq.~(\ref{eq:dyanmicalEquationModes})
\begin{equation}\label{eq:langevinEquation}
B(Q,m) \dot{u}_{qm} = -A(Q,m) u_{qm} + \xi^{\rm th}_{qm} +\xi^{\rm ac}_{qm},
\end{equation}
where $\xi^{\rm th}_{qm}$ and $\xi^{\rm ac}_{qm}$ denote the passive (thermal) and active forces 
respectively. The statistical properties of the thermal noise are given in the standard way
\begin{equation}
\label{eq:thermalNoise}
\langle\xi^{\rm th}_{qm}(t)\rangle = 0,
\end{equation}
\begin{equation}
\langle\xi^{\rm th}_{qm}(t)\xi^{\rm th *}_{q'm'}(t')\rangle = 
\frac{2 k_{\rm B}T}{\kappa}B(Q,m)\delta_{qq'}\delta_{mm'}\delta(t-t'),
\end{equation}
where $k_{\rm B}$ is Boltzmann constant, $T$ is the temperature and the star, $(^{*})$, denotes the complex conjugate.

For the active fluctuations, we write
\begin{equation}
\langle\xi^{\rm ac}_{qm}(t)\rangle = 0,
\end{equation}
\begin{equation}
\langle\xi^{\rm ac}_{qm}(t)\xi^{\rm ac *}_{q'm'}(t')\rangle = 
\frac{[F(Q,m)]^2}{2\tau_{\rm ac}}e^{-|t-t'|/\tau_{\rm ac}}
\delta_{qq'}\delta_{mm'},
\end{equation}
where $\tau_{\rm ac}$ is the correlation time of the active forces, and the physics of the active processes 
will be captured in our choice of active force density, $F(Q,m)$~\cite{gov_membrane_2004}. 
We will consider only the simplest case where direct forces acting on the membrane is 
constant, i.e., $F(Q,m)=F$, although more realistic models could be 
considered~\cite{turlier_equilibrium_2016,manneville_active_2001}.

\subsection{Thermal fluctuations}

First we consider the case when there are no active fluctuations, i.e., $\xi^{\rm ac}_{qm}=0$. 
Solving Eq.~(\ref{eq:langevinEquation}) by Fourier transform in time (assuming any initial conditions have decayed away) 
yields the following covariance
\begin{align}
\langle u_{qm}\left(t\right) u^*_{q'm'}\left(t'\right)\rangle^{\rm th}
= \frac{k_{\rm B}T}{\kappa A}e^{-\lambda(Q,m) |t-t'|}\delta_{qq'}\delta_{mm'}.
\end{align}
where ${}^{\rm th}$  identifies these as thermal (passive) fluctuations. If we Fourier transform in time with the convention 
$u(t)=\int \mathrm{d} \omega/(2\pi) \, u (\omega)e^{i \omega t}$ we can also find the frequency domain covariance
\begin{equation}\label{eq:thermalFreqCov}
\langle u_{qm}(\omega) u^{*}_{q'm'}(\omega')\rangle^{\rm th}
 = \frac{2k_{\rm B} T B \delta_{qq'}\delta_{mm'}\delta(\omega-\omega')}
{\kappa(A^2+B^2\omega^2 )}
\end{equation}
which we will make use of later.

\begin{figure}
\includegraphics[width=0.45\textwidth]{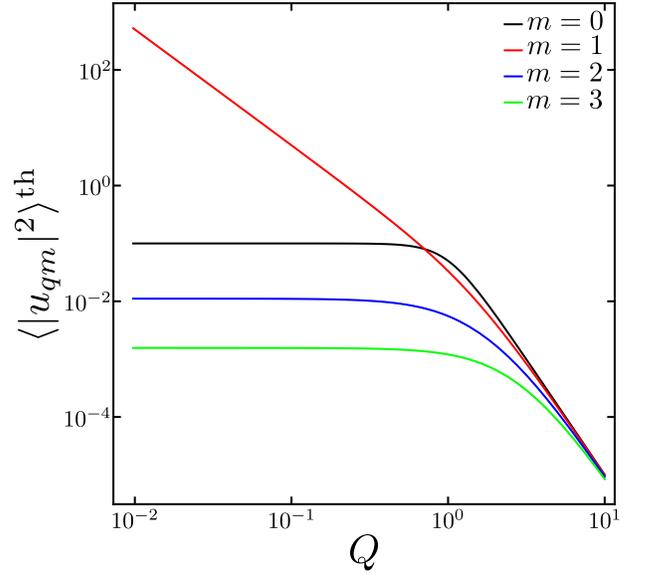}
\caption{Equal time variance for thermal fluctuations as a function of dimensionless wavenumber $Q$ 
for modes $m=0,1,2,3$. 
The dimensionless surface tension is $\Sigma=1/2$ and the bending rigidity is $\kappa/(k_{\rm B}T)=10$.}
\label{fig:thermalFluctuations}
\end{figure}

Then the equal time variance is given by
\begin{align}
\langle| u_{qm}|^2\rangle^{\rm th}
&= \frac{k_{\rm B}T}{\kappa}\bigg[(Q^2+m^2)^2 - \frac{1}{2}Q^2 -\frac{5}{2}m^2
\nonumber \\
&+\frac{3}{2}-\Sigma(1-Q^2-m^2)\bigg]^{-1},
\end{align}
in accordance with the equipartition theorem~\cite{fournier_critical_2007}. 
The equal time covariance is plotted against $Q$ in Fig.~\ref{fig:thermalFluctuations} for $m=0, 1, 2, 3$.
Here the dimensionless surface tension is chosen to be $\Sigma=1/2$ and we choose a typical order of magnitude for the bending rigidity, $\kappa/(k_{\rm B}T)=10$~\cite{rautu_role_2017}.

A striking prediction is the divergence of the $m=\pm 1$ modes, i.e., criticality, with 
a power-law scaling $\langle| u_{q1}|^2\rangle^{\rm th}\sim Q^{-2}$, in the limit $Q \to 0$.
This criticality is due to the $m=\pm 1$ modes being one-dimensional Goldstone modes in the long wavelength limit.
In other words, for small $Q$, they only locally translate the cross-section of the tube which does not alter the energy of the tube. 
The equilibrium properties of such fluctuations, such as excess area and length fluctuations, are discussed in 
Ref.~\cite{fournier_critical_2007}. 
Due to the one-dimensional character of these modes, it is expected that the criticality will be preserved even in 
the anharmonic regime~\cite{fournier_critical_2007}.

\subsection{Active fluctuations}

Turning our attention to the case of active fluctuations, we will find the statistical properties of the shape fluctuations 
due to purely active noise. 
We assume that the active and thermal noise terms are uncorrelated, hence the total shape fluctuations can be found 
by simply adding the active and passive contributions.

\begin{figure}
\includegraphics[width=0.45\textwidth]{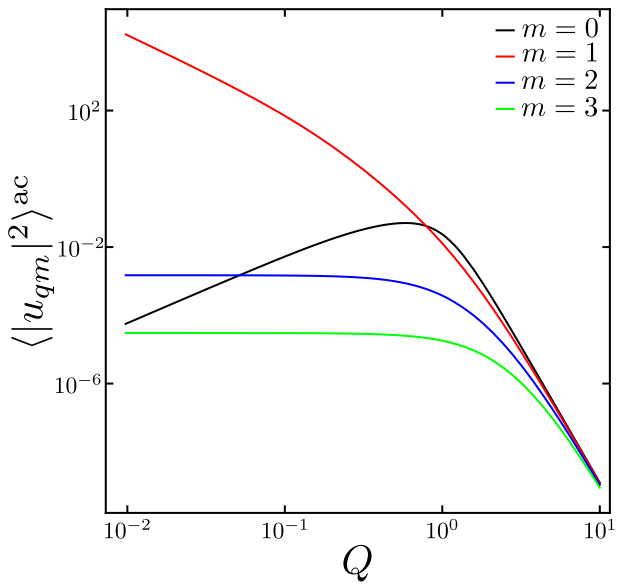}
\caption{Equal time variance for active fluctuations of the membrane tube as a function of dimensionless 
wavenumber $Q$ for modes $m=0,1,2,3$. 
The dimensionless parameters are $\Sigma=1/2$, $\chi=0$, $F^2=2.5$ and $\tau_{\rm ac}/\tau =10$.}
\label{fig:activeFluctuations}
\end{figure}

To find the covariance, we Fourier transform in time to find
\begin{equation}\label{eq:activeFreqCov}
\langle u_{qm}(\omega) u^{*}_{q'm'}(\omega')\rangle^{\rm ac}
 = \frac{F^2 \delta_{qq'}\delta_{mm'}\delta(\omega-\omega')}
{(A^2+B^2\omega^2 )(1+\tau_{\rm ac}^2\omega^2 )}.
\end{equation}
Inverting the Fourier transform for $\omega$ and $\omega'$ gives the covariance in time, which after some algebra gives
\begin{align}
& \langle u_{qm}\left(t\right) u_{q'm'}^{*}\left(t'\right)\rangle^{\rm ac} 
\nonumber \\
& = \frac{F^2}{2A} 
\frac{A\tau_{\rm ac} e^{-|t - t'|/\tau_{\rm ac}} - Be^{-(A/B) |t - t'| }}{A^2\tau_{\rm ac}^2-B^2}
\delta_{qq'}\delta_{mm'}. 
\end{align}
Hence the equal time variance becomes
\begin{equation}\label{eq:activeVariance}
\langle| u_{qm}|^2\rangle^{\rm ac} = 
\frac{F^2}{2A(A\tau_{\rm ac}+B)}.
\end{equation}

As this quantity depends on the dissipation in the system through the factor $B$, it is obvious that the 
fluctuations are nonequilibrium. 
If we assume that the activity correlation time is an order of magnitude more than the viscous time scale, 
$\tau_{\rm ac}/\tau \approx 10$, and that the forces exerted to the membrane is about 
$1$~pN over an area $r_0^2$, then one can estimate $F\sim 1$--$10$~\cite{sachin_krishnan_thermal_2018}. 
Using these parameters along with $\chi=0$, $F^2=2.5$ we plot in Fig.~\ref{fig:activeFluctuations}
the active fluctuations given by Eq.~(\ref{eq:activeVariance}). A peak in the $m=0$ mode is apparent, the position of which depends on the relative value of the active time-scale 
$\tau_{\rm ac}/\tau$.
The decay in active fluctuations of the $m=0$ mode as $Q\to 0$ is due to the viscous damping that suppresses such non-equilibrium fluctuations. This does not appear in the thermal case as the thermal force fluctuation scales like $B(Q,m)$.

The divergence at small $Q$ observed in the $m=1$ modes is retained but with an exponent that differs from the thermal case according to $\langle| u_{q1}|^2\rangle^{\rm ac}\sim Q^{-2}\log Q$. It is interesting to briefly consider the more general case when $F$ is not constant, and its effect on the critical nature of the $m=\pm 1$ mode. In particular if, say by a curvature sensitive coupling, $F(Q,m)\sim Q^{n/2}$ in the small $Q$ limit, then the fluctuations would scale as $\langle| u_{q1}|^2\rangle^{\rm ac}\sim Q^{n-2}\log Q$. For example, if the active noise is coupled to the change in curvature from the ground state of the tube, then one might expect $F(Q,m)\sim 1/(1-m^2-Q^2)$, or $n=-4$ (see Appendix \ref{app:curvature}), which leads to bending mode fluctuations that scale like $\langle| u_{q1}|^2\rangle^{\rm ac}\sim Q^{-6}\log Q$. 

As the $m=\pm 1$ mode is the softest on the tube, it dominates the real space fluctuations and, by Parseval's Theorem, we have $\langle |u(\theta,z)|^2\rangle \approx 2\int_{2\pi/L}^{\infty} \mathrm{d}q/(2\pi)\langle| u_{q1}|^2\rangle$. This implies that by measuring the real space fluctuations of the tube, either through florescence microscopy or other indirect means \cite{valentino_fluctuations_2016,georgiades_flexibility_2017}, and then varying the length of the tube, and thus the cut-off wavelength, one could infer the long-wavelength form of the active noise. This experimental assay would provide a simple test of our predictions and might help to differentiate between different types of active noise. A similar method has been used to quantify active noise in neurites~\cite{garatel_time-resolved_2015}.

\begin{figure}
\includegraphics[width=0.45\textwidth]{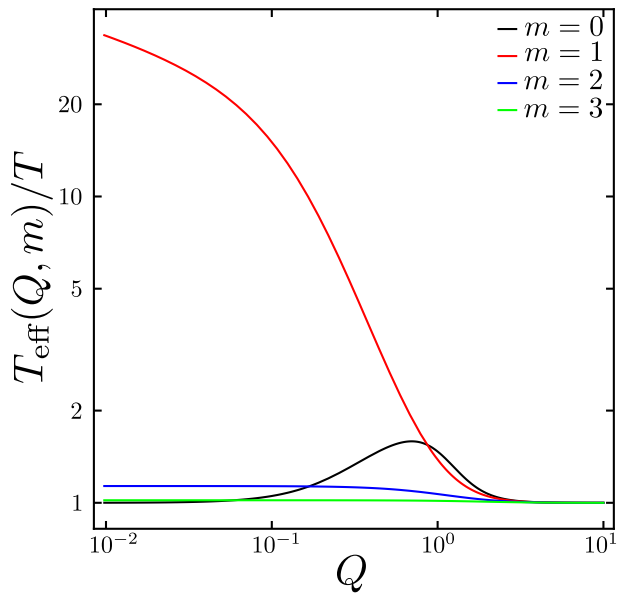}
\caption{Effective temperature of the membrane tube with both thermal and active fluctuations, Eq.~(\ref{eq:effectiveTemp}), plotted as a function of dimensionless wavenumber $Q$ for modes $m=0,1,2,3$. 
The dimensionless parameters are $\Sigma=1/2$, $\kappa/(k_{\rm B}T)=10$, $\chi=0$, $F^2=2.5$, and $\tau_{\rm ac}/\tau =10$.}
\label{fig:effectiveTemp}
\end{figure}

Another possible observable is the effective temperature as a function of Fourier parameters:
\begin{equation}\label{eq:effectiveTemp}
\frac{T_{\text{eff}}\left(Q,m\right)}{T} = 1 + \frac{\langle| u_{qm}|^2\rangle^{\rm ac}}
{\langle| u_{qm}|^2\rangle^{\rm th}}.
\end{equation}
This quantity is plotted in Fig.~\ref{fig:effectiveTemp} for the same parameters of $F^2=2.5$, $\tau_{\rm ac}/\tau=10$, 
$\chi=0$, $\kappa/(k_{\rm B}T)=10$ and $\Sigma=1/2$. 
The plot shows that, for long tubes, the highest effective temperature is found in the $m=1$ modes and that these 
are likely to dominate the spectrum.

\begin{figure}
\includegraphics[width=0.45\textwidth]{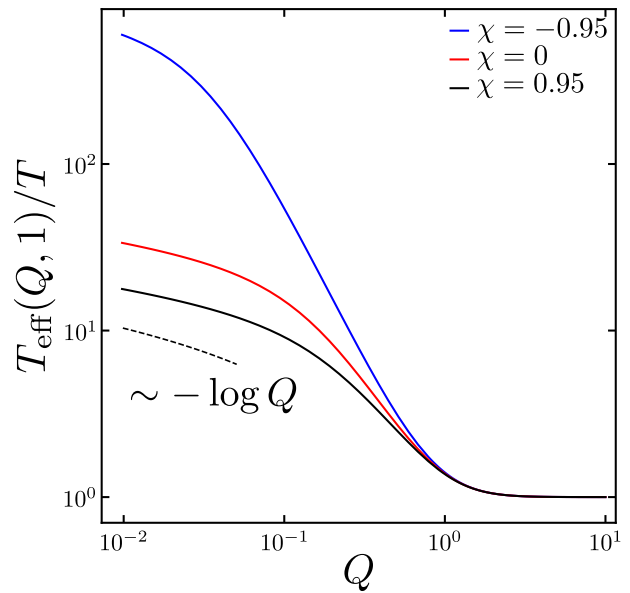}
\caption{Effective temperature of the membrane tube with both thermal and active fluctuations, Eq.~(\ref{eq:effectiveTemp}), plotted as a function of dimensionless wavenumber $Q$ for modes $m=1$ for different values of relative viscosity $\chi$. 
The dimensionless parameters are $\Sigma=1/2$, $\kappa/(k_{\rm B}T)=10$, $F^2=2.5$, and $\tau_{\rm ac}/\tau =10$.
The dotted line shows the asymptotic approximation obtained by using a Zimm model for a rod in a viscous fluid.}
\label{fig:effectiveTempViscous}
\end{figure}

Measuring the temperature of fluctuations of long tubes, for example those pulled from GUVs~\cite{valentino_fluctuations_2016}, 
and varying the viscosity of the exterior fluid may provide a way to quantify the magnitude and time constant of such active 
correlations in experiment. 
Figure~\ref{fig:effectiveTempViscous} shows the effective temperature of the $m=\pm 1$ modes for varying relative viscosity 
$\chi$, along with the asymptotic result predicted using a Zimm model for such modes (see Appendix \ref{app:zim}). As the $m=\pm 1$ modes dominate much of the fluctuation spectrum, and given the small size of most tubes formed in real membrane systems, it might prove difficult to resolve the other modes directly. Hence the $m=\pm1$ modes are probably the best candidate for such a direct measurement with varying viscosities.

\begin{figure}
\includegraphics[width=0.45\textwidth]{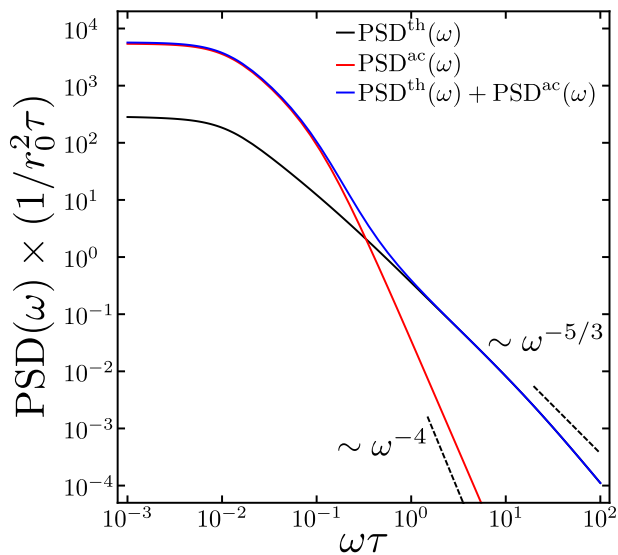}
\caption{The Power Spectral Density, $\text{PSD}(\omega)$, plotted against frequency, $\omega$, in the thermal, active and combined case. The dimensionless parameters are $\Sigma=1/2$, $\kappa/(k_{\rm B}T)=10$, $\chi=0$, $F^2=2.5$, and $\tau_{\rm ac}/\tau =10$. We choose a tube of length $L=100 r_0$.}
\label{fig:PSDPlot}
\end{figure}

Finally, we compute the Power Spectral Density (PSD) for the active and thermal cases from the frequency domain correlation functions, $\langle|u_{qm}(\omega)|^2\rangle^{\rm th, \rm ac}$, given by Eqs.~(\ref{eq:thermalFreqCov}) and (\ref{eq:activeFreqCov}). The PSD is given by
\begin{equation}
\text{PSD}^{\rm th, \rm ac}(\omega)=2\sum_{Q>Q_{\text{min}},m\geq 0}\langle|u_{qm}(\omega)|^2\rangle^{\rm th, \rm ac}
\end{equation}
where the factor of $2$ comes from counting negative $q$ and $m$ modes, and $Q_{\text{min}}$ is the cutoff set by the length of the tube. We plot the $\text{PSD}$ in Fig.~\ref{fig:PSDPlot} for the same parameters as before and using discrete values of $Q=2\pi n r_0/L$ to preform the summation where we choose $n\in[1,50]$, $m\in[0,3]$ and the tube length $L=100r_0$.

The combined PSD, $\text{PSD}^{\rm th}(\omega)+\text{PSD}^{\rm ac}(\omega)$, displays three regimes; a plateau at small $\omega$ mostly dominated by active noise, a transitional region governed by the active noise which scales like $\text{PSD}(\omega)\sim \omega^{-4}$, and finally a high frequency regime dominated by the thermal behaviour. This final regime scales in a similar way to that for a flat membrane, $\text{PSD}(\omega)\sim\omega^{-5/3}$~\cite{zilman_undulations_1996,zilman_membrane_2002}. Crucially, the low frequency regime, which is the only region that can be probed by current experimental techniques, shows a dramatic difference between the passive and active cases. This could be measured directly using current experimental methods~\cite{valentino_fluctuations_2016,allard_fluctuations_2020}, using an ATP-depleated system as a reference for the passive thermal case. A similar technique has been used to analyse the active ``flicker'' of red blood cells~\cite{garcia_direct_2015}, and could provide a direct way of quantifying active behaviour in membrane tubes. Varying the external viscosity could also provide an additional probe of the form of the active forcing.

\section{Summary and discussion}
\label{sec:discussion}

In this paper, we have investigated the passive and active dynamics of a fluid membrane tubes. 
Using Onsager's variational formalism, we have calculated the full relaxation dynamics for the Fourier modes in the 
shape of the membrane tube, assuming a small deformation limit.
This work accounts for variations in surface tension through the use of the Lagrange multiplier, $\zeta$, imposing local incompressibility, previously only considered in the axis-symmetric 
case~\cite{boedec_pearling_2014}. We also consider the viscosity contrast between the interior and exterior ambient fluid.

The scaling behaviour of the relaxation modes is analysed and characterised in both the long and short wavelength limits. 
In the short wavelength limit, we recover the scaling behaviour of a flat membrane for all angular modes. 
More interesting behaviour is found in the long wavelength limit, particularly in the case of the bending modes 
($m=\pm 1$), where we find a relaxation rate that scales like that of the normal modes of an elastic rod in a viscous fluid. 
We also reproduce the pearling instability growth rate found by Boedec \textit{et.\ al.}~\cite{boedec_pearling_2014}, 
which is recovered when we set $m=0$ and choose a sufficiently high value of the surface tension 
($\Sigma>3/2$). 
These relaxation dynamics are significantly different from those found for flat 
membranes~\cite{fournier_hydrodynamics_2015,seifert_hydrodynamics_1994} or for 
spherical vesicles~\cite{sachin_krishnan_relaxation_2016}. 
In the case of the spherical vesicles, the system can be written purely in terms of one Fourier mode due to the coupling 
imposed by spherical symmetry. 
This does not happen in the case of the tube as $Q$ and $m$ are independent of one another.

We then make use of these relaxation equations to compute the fluctuation spectra for passive thermal fluctuations and 
a simple minimal model of active fluctuations~\cite{sachin_krishnan_thermal_2018,gov_membrane_2004}. 
The active noise breaks the fluctuation dissipation theorem, as is apparent in the presence of dissipative terms in the mean
square fluctuations, see Eq.~(\ref{eq:activeVariance}). 
The active noise for a ``direct force'' term also shows a modified critical behaviour of the bending modes ($m=\pm 1$) in the long-wavelength limit found in thermal fluctuations~\cite{fournier_critical_2007}. It is likely that more complex curvature couplings in the long-wavelength limit could also modify the critical exponent. We have proposed a possible experimental assay based on measuring the real space fluctuations of the tube and varying its length. This could be used to infer the functional form of the active noise experimentally in both \textit{in-vivo} and \textit{in-vitro} systems \cite{valentino_fluctuations_2016,georgiades_flexibility_2017,nixon-abell_increased_2016}. We also compute the effective temperature of the system with both thermal and active fluctuations and show that, for long tubes, 
the clearest signature of the active noise is in the bending modes. This should be a measurable prediction with current experimental setups, \textit{e.g.}, using a similar approach used by Valentino \textit{et.~al.}~\cite{valentino_fluctuations_2016} and changing the external viscosity. 

Finally we compute the Power Spectral Density for thermal, active and combined fluctuations of membrane tubes, a quantity that is directly measurable with current experimental setups using optical tweezers \cite{valentino_fluctuations_2016,allard_fluctuations_2020}, and show that the active fluctuations dominate the measurable low frequency regime.
This could be used to directly infer information about the size of forces and activity time scales for different sources of activity. It could also be interesting to consider the effect of different forms of the active force, $F(Q,m)$, particularly in the case where this includes a length-scale associated to curvature coupling (as might be the case with proteins changing conformation). 

In this work we have assumed that the bilayer is symmetric, such that there is no spontaneous curvature term in the Helfrich-Canham energy. However if the tubes were formed by a coat of proteins with preferred curvature (BAR-domain proteins, for example), then an additional spontaneous curvature term would appear in the elastic free energy~\cite{kabaso_on_2012,iglic_curvature-induced_2006}. This is known to have an effect on the relaxation rates and criterion for the pearling instability~\cite{jeleri_pearling_2015}.

More complex boundary conditions allowing for slip and permeation of the membrane with the ambient fluid could also be considered by adding the corresponding dissipative terms to the Rayleighian~\cite{yasuda_dynamics_2018,manneville_active_2001}. These effects would be most relevant at very short wavelengths and thus only modify the high $(Q,m)$ part of the relaxation rates and fluctuation spectra~\cite{yasuda_dynamics_2018}. Permeation may also play a role for the $m=0$ mode at very long wavelengths and thus be important in tubes undergoing volume changes, by osmotic swelling for example \cite{al-izzi_hydro-osmotic_2018}. It is also possible that in some scenarios the membrane may allow slip freely but support no shear stress across the membrane. In this case a change in boundary condition from no-slip to no-shear-stress would need to be considered~\cite{shlomovitz_membrane-mediated_2011}.

Perhaps the most pressing open question in the field of active membranes is what functional form is best used to represent 
the active fluctuations and if all current descriptions can be unified in some manner. 
The simple model of a direct force used in this paper has been 
used successfully throughout the literature to describe real systems~\cite{gov_membrane_2004,garcia_direct_2015,prost_shape_1996,sachin_krishnan_thermal_2018}. However, it is not clear how well motivated this is at the microscopic level. The model we have used involves the assumption that the work done by the active forces per unit time is constant, although another possible coarse-grained model might assume constant applied force~\cite{gov_membrane_2004}.
More complex models of activity have been proposed for specific situations, for example using dipole forces and allowing fluid 
permeation of the membrane~\cite{manneville_active_2001}.
However, a general framework is lacking and the effect on observable phenomena is not yet well understood.

For future work, it would be interesting to consider the effects of different formulations of activity, both in tubes and other geometries. An example might be  active pumps that transport ions across the membrane, and thus increasing the tube's osmotic pressure \cite{al-izzi_hydro-osmotic_2018}. A further study could focus on coupling the active fluctuations to a field associated with the density of some spontaneous curvature-inducing proteins on the membrane. This has already been studied for a flat membrane in Ref.~\cite{ramaswamy_nonequilibrium_2000} where a curvature induced instability was shown to arise. The interplay between this instability and the highly curved geometry of a membrane tube could be quite rich. 
It may also be of interest to consider the effect of a viscoelastic ambient fluid, as this may give a better approximation to the 
cytoplasm in cells. 
Not only would this give potentially richer dynamics, due to the presence of of an additional time scale, but it could also 
be useful in understanding more realistic biological 
processes~\cite{komura_dynamics_2015,nixon-abell_increased_2016,abounit_wiring_2012}.

\acknowledgements

S.C.A.-I.\ would like to acknowledge funding from the following sources; the EPSRC under grant number EP/L015374/1 CDT in Mathematics for Real-World Systems, the Labex CelTisPhyBio (ANR-11-LABX-0038, ANR-10-IDEX-0001-02), the EMBL-Australia program, and Tokyo Metropolitan University under the Graduate Short-term Inbound and 
Outbound Program. S.C.A.-I. would also like to thank Prof.~S.~Komura for his generous hospitality during his stay at TMU. M.S.T. acknowledges the generous support of the Japan Society for the Promotion of Science (JSPS), via a long term fellowship, and the hospitality of both Prof.~R.~Yamamoto (Kyoto University) and Prof.~S.~Komura at Tokyo Metropolitan University. S.K.\ acknowledges support by a Grant-in-Aid for Scientific Research (C) (Grant No.\ 18K03567 and Grant No.\ 19K03765) from the JSPS, and support by a Grant-in-Aid for Scientific Research on Innovative Areas ``Information Physics of Living Matters'' (Grant No.\ 20H05538) from the Ministry of Education, Culture, Sports, Science and Technology of Japan.

\appendix
\section{Expressions for $\Phi^\pm_{qm}$, $\Psi^\pm_{qm}$, $\Xi^\pm_{qm}$}
\label{app:coefficients}

Here we give expressions for the scalar Laplace function decompositions for the Stokes equations after imposing the 
boundary condition, Eq.~(\ref{eq:boundaryCondition}), and making use of the continuity equation to eliminate 
$v^\theta_{qm}$. 
This gives
\begin{widetext}
\begin{align}
\Phi^+_{qm} & = 2r_0\bigg[Q \left(4i v^z_{qm} - 3Q r_0\dot{u}_{qm}\right)K_{m-1}^2 - 4 m \left(Q r_0 \dot{u}_{qm} -2i v^z_{qm}\right)K_{m-1}K_m+4\left( r_0 \dot{u}_{qm} +i v^z_{qm}\right)K_m^2
\nonumber \\ 
&-Q^2 r_0 \dot{u}_{qm} K_{m+1}^2\bigg]\times\bigg[7 Q^3K_{m-1}^3 +2\left(9m-8\right)Q^2 K_{m-1}^2K_m +4 Q\left(m\left(m-8\right)-2Q^2\right)K_{m-1}K_m^2
\nonumber \\
&- 8 m\left(m^2+Q^2\right)K_m^3 +Q^3K_{m+1}^3 \bigg]^{-1},
\end{align}
\begin{align}
\Psi^+_{qm} & = r_0\bigg[-8 K_m^2 \left(v^z_{qm} \left(m^4+2 (m+1) m Q^2 + Q^4 \right)- i Q r_0 \dot{u}_{qm} \left(m (3 m+2)+Q^2 \right) \right)
\nonumber \\ 
&+ 8 Q K_{m-1} K_m \left(\left(m^3 + (m-2) Q^2\right) v^z_{qm}  - i (m-2) Q r_0 \dot{u}_{qm} \right)
\nonumber \\ 
&+2 Q^2 \left(3 K_{m-1}^2+K_{m+1}^2\right) \left(\left(m^2 + Q^2 \right) v^z_{qm} -iQr_0 \dot{u}_{qm} \right)\bigg]\times \bigg[ m Q \Big(8 \left(m^3+m Q^2 \right) K_m^3 - 7 Q^3 K_{m-1}^3 
\nonumber \\
&- Q^3 K_{m+1}^3 + 2 (8-9 m) Q^2 K_m K_{m-1}^2 + 4 Q \left(2 Q^2 -(m-8) m\right) K_m^2 K_{m-1}\Big)\bigg]^{-1},
\end{align}
\begin{align}
\Xi^+_{qm} & = r_0\bigg[-8 i K_m^2 \left(v^z_{qm} \left(m^2+Q^2 \right)-i (m+1) Q r_0 \dot{u}_{qm}\right)+\left(6 i Q^2 K_{m-1}^2 + 2 i Q^2 K_{m+1}^2\right) v^z_{qm}
\nonumber \\ 
&- 8 Q K_m K_{m-1} \left(Q r_0^2 \dot{u}_{qm}-i m v^z_{qm}\right)\bigg]\times 
\bigg[ Q \Big(-8 \left(m^3+m Q^2 \right) K_m^3 + 7 Q^3 K_{m-1}^3 + Q^3 K_{m+1}^3
\nonumber \\
&+ 2 (9 m-8) Q^2 K_m K_{m-1}^2 + 4 Q \left((m-8) m-2 Q^2\right) K_m^2 K_{m-1}\Big)\bigg]^{-1},
\end{align}
\begin{align}
\Phi^-_{qm}&  = r_0\bigg[ I_m^2 \left(\left(m^2-1\right) r_0\dot{u}_{qm}- i Q v^z_{qm} \right)+Q I_{m-1}^2 \left(Q r_0 \partial u_{qm} - i v^z_{qm} \right) - 2 m I_m I_{m-1} \left(Q r_0 \dot{u}_{qm} - i v^z_{qm}\right)\bigg]
\nonumber \\
&\times \bigg[ Q \left(Q^2 I_{m-1}^3 + \left(2 (m-2) m-Q^2 \right) I_m^2 I_{m-1}+(2-3 m) Q I_m I_{m-1}^2 + m Q I_m^3\right) \bigg]^{-1},
\end{align}
\begin{align}
\Psi^-_{qm} & = r_0\bigg[ -2 I_m I_{m-1} \left(m^3 v^z_{qm} + (m-1) Q^2 v^z_{qm} - i (m-1) Q r_0 \dot{u}_{qm} \right)
\nonumber \\
&+ Q I_{m-1}^2 \left(\left(m^2  + Q^2\right) v^z_{qm} - Qi r_0 \dot{u}_{qm}) \right)
\nonumber \\
&+ i I_m^2 \left(r_0 \dot{u}_{qm} \left(2 m (m+1)+Q^2 \right) + i Q v^z_{qm} \left(m (m+2) + Q^2 \right)\right)\bigg]
\nonumber \\
&\times \bigg[ m Q \left(Q^2 I_{m-1}^3 + \left(2 (m-2) m-Q^2 \right) I_m^2 I_{m-1} + (2-3 m) Q I_m I_{m-1}^2 + m Q I_m^3\right)\bigg]^{-1},
\end{align}
\begin{align}
\Xi^-_{qm} & = r_0\bigg[ I_m I_{m-1} \left(-Q r_0^2 \dot{u}_{qm} + 2 i m v^z_{qm} \right) +  I_m^2 \left(\left(m +1\right)r_0
\dot{u}_{qm} + i Q v^z_{qm} \right)-i Q v^z_{qm} I_{m-1}^2\bigg]
\nonumber \\
&\times \bigg[ Q \left(Q^2 I_{m-1}^3 + \left(2 (m-2) m- Q^2 \right) I_m^2 I_{m-1} + (2-3 m) Q I_m I_{m-1}^2+m Q I_m^3\right) \bigg]^{-1},
\end{align}
\end{widetext}
where the modified Bessel functions $K_m$ and $I_m$ are evaluated at $r=r_0$.

\section{Relaxation dynamics of linear Zimm model}
\label{app:zim}

Here we consider the relaxation dynamics of small planar normal perturbations to a thin elastic rod whose position is given by
\begin{equation}
\vec{r} = \left(x(t) \cos q z,0,z\right)\text{,}
\end{equation}
and has geodesic curvature $k_{\rm g}=-q^2 x \cos q z$. We are motivated to study the relaxation dynamics of this system as, on length scales much larger than the tube radius, the $m=1$ mode of a membrane tube can likely be thought of as the dynamics of a thin elastic rod. Here membrane flows are small and most of the friction comes from the drag of the tube through the bulk fluid. We want to see if, at least at a scaling level, this can give a simple understanding of the long wavelength relaxation dynamics.

If we assume $q x(t)\ll 1$, the elastic force per unit length on the rod is given by
\begin{align}
\vec{f} &= \left(-K \nabla^2 k_{\rm g} + S k_{\rm g},0,0\right)
\nonumber \\ 
&= \left(-K q^4 x \cos q z -S q^2 x \cos q z,0,0\right),
\end{align}
where $K$ is the bending rigidity of the rod and $S$ is the tension~\cite{audoly_elasticity_2010}.

We can write the dynamics of this rod as a continuous Zimm model 
\begin{equation}
\dot{\vec{r}} = - \int \mathrm{d}\vec{s} \, \Lambda\left(\vec{r}-\vec{s}\right) \vec{f}\left(\vec{s}\right),
\end{equation} 
where $\Lambda\left(\vec{r}-\vec{r}'\right)$ is the Oseen tensor~\cite{doi_theory_1986}
\begin{equation}
\Lambda\left(\vec{r}-\vec{s}\right) = \frac{1}{8\pi\eta|\vec{r}-\vec{s}|}
\left[\mathbb{I}-\frac{\left(\vec{r}-\vec{s}\right)\otimes \left(\vec{r}-\vec{s}\right)}{|\vec{r}-\vec{s}|^2}\right].
\end{equation}
Here $\otimes$ is the tensor product and $\mathbb{I}=\delta_{\alpha\beta}\vec{e}_{\alpha}\otimes\vec{e}_{\beta}$.

At linear order and in the long-wavelength limit,  this gives
\begin{equation}
\dot{x} \approx \frac{(K q^4 +Sq^2)\text{Ci}(q r_0)}{4\pi\eta}x,
\end{equation}
where $\text{Ci}\left(q r_0\right)=-\int_{r_0}^\infty \mathrm{d}x \, \cos (q x)/x$ is the cosine integral function, and 
we have chosen a short-wavelength cutoff of the rod radius, $r_0$.
Then we obtain a relaxation rate that scales as $\lambda \sim -\left(\gamma +\log Q\right)Q^2$ in the small $Q$ 
limit, where $\gamma$ is the Euler-Mascheroni constant. 
This result agrees, at the scaling level, with relaxation dynamics of a membrane tubes bending mode in the long wavelength limit. It also gives the correct scaling for the effective temperature of fluctuations in the long wavelength limit, dotted line in Fig.~\ref{fig:effectiveTempViscous}.

\section{Possible direct curvature coupling}\label{app:curvature}
Here we briefly outline a possible scheme to give active fluctuations with direct curvature coupling. If instead of writing the shape equation in terms of the local deviation from the equilibrium radius, $r_0$, we write it as a local deviation from the equilibrium mean curvature, $H=1/(2 r_0)$. The we can write an equation of the form
\begin{equation}
\delta \dot H_{qm} B(Q,m) = - A(Q,m)\delta H_{qm} + \xi^{\text{ac}, H}_{qm} 
\end{equation}
where $\delta H = (1-m^2-Q^2)u_{qm}$ and $\xi^{\text{ac},H}_{qm}$ gives the active direct curvature coupling noise with the following statistical properties
\begin{align}
&\langle\xi^{\text{ac},H}_{qm}(t)\rangle = 0\\
&\langle\xi^{\text{ac},H}_{qm}(t)\xi^{\text{ac},H*}_{q'm'}(t')\rangle=\frac{f^2}{2\tau_{\rm ac}}e^{-|t-t'|/\tau_{\rm ac}}
\delta_{qq'}\delta_{mm'},
\end{align}
where $f$ is a constant. In the shape equation for $u_{qm}$, Eq.~(\ref{eq:langevinEquation}), this corresponds to $F(Q,m)=f/(1-m^2-Q^2)$ which is discussed briefly in the main text.


\end{document}